# Power Control for Harmonic Utility in Non-Orthogonal Multiple Access based Visible Light Communications

Quoc-Viet Pham and Choong Seon Hong
Department of Computer Science and Engineering, Kyung Hee University
vietpq90@khu.ac.kr, cshong@khu.ac.kr

**Abstract**

Visible light communication is a promising technique in future networks due to its advantages of high data rate and licensed-free spectrum. In addition, non-orthogonal multiple access (NOMA) is considered as a candidate of multiple access schemes in 5G networks and beyond. In this paper, we study the power control problem in NOMA-based visible light communication. An optimization problem is formulated, where the objective is the harmonic-rate utility function. The optimal solution is achieved optimally by equivalently transforming the underlying problem into a convex problem. Simulation result is provided to validate performance of the proposed approach.

## 1. Introduction

Due to the global demand for high data-rate applications in wireless networks and recent advances in solid-state lighting, visible light communication (VLC) has been considered as a potential and effective communication paradigm. Using light-emitting diodes (LEDs) for VLC has been received much attention from both industry and academia due to lots of offered advantages [1], for example, easy installation of lighting equipment with LEDs, possibility of high data rate communication, and low cost, low power consumption and long life expectancy. However, the main challenge of VLC systems is the narrow modulation bandwidth of the LEDs (in megahertz range), which strictly limits the achievement of high data rate.

Non-orthogonal multiple access (NOMA) is a promising candidate among multiple access schemes in 5G wireless networks. As opposed to traditional multiple access, for example, TDMA, FDMA, and OFDMA, where each radio resource is exclusively assigned to at most one user, NOMA enables simultaneous transmission of multiple users by adopting the superposition technique at the transmitter side and successive interference cancelation at the receiver side. Integration of NOMA in VLC systems is reasonable thanks to some reasons [2, 3], for example, NOMA is good at high SNR, which is typical to VLC systems due to short distance between the LED and users and line-of-sight (LOS) path, the channels remain almost constant in VLC systems, which is vital to NOMA's functionalities in decoding order and power allocation, and the LED transmitter can be served as a small access point, whose purpose is to provide services to a small number users, while NOMA is to multiplex a few number of users (often two users to reduce complexity at the user side).

In NOMA, fairness and performance can be improved and supported through an appropriate power allocation [4]. Some recent literature have devoted to power allocation in VLC systems. In [5], the effect of fixed power allocation on the system coverage probability and system ergodic capacity was studied in a single-LED VLC system. In [2], a channel dependent power allocation scheme, named Gain Ratio Power Allocation (GRPA), was proposed to ensure fairness among users since a smaller transmit power is sufficient for a user with higher channel gain to decode signals of users with lower channel gains and a user with lower channel gain needs higher transmit power to compensate the large interference from users with higher channel gains. Exploiting the feasibility of precise positioning in VLC networks, a user group method to reduce inter-cell interference was proposed in [6]. Additionally, two power allocation schemes, total sum rate and max-min rate, were proposed to guarantee quality of service (QoS) of users.

Actually, there is a tradeoff between the fairness and sum rate of users. This tradeoff can be presented by using the $\alpha$-fair utility function and well studied in literature [4]. In this paper, we study a special case of the fairness index $\alpha = 2$, harmonic-rate utility. Although the original problem is non-convex, we show that it can be equivalently transformed into a convex problem and then can be solved optimally by interior-point methods to achieve the optimal solution. Simulation result is conducted to show performance of the proposed algorithm.

## 2. System Model and Formulation

Consider the downlink of a VLC network with one LED transmitter an $M$ users. NOMA allows the LED transmitter simultaneously transmits the data to $M$ users in the same frequency radio resource at the same time. In addition, in NOMA, the superposition coding technique is deployed at the transmitter while multi-user detection algorithms such as successive interference cancellation (SIC) are used to detect the desired signals at the receiver side.

For the channel model, we consider the LOS propagation path and use the generalized Lambertian emission model, where the channel DC gain is proportional to the inverse of the squared distance between the LED transmitter and user [Eq. (10), 7], Specifically, the DC channel gain of user $m$ is expressed by

$$h_m = \frac{A_m}{d_m^2} R_0(\varphi_m) T_s(\phi_m) g(\phi_m) \cos\phi_m, \qquad (3)$$

for $0 \leq \phi_m \leq \Phi_m$ and $h_m = 0$ otherwise, where $\Phi_m$ is the field of view (FOV) of user $m$, $\varphi_m$ and $\phi_m$ is the angle of irradiance and the angle of incidence, respectively. In Eq. (3), $A_m$ is the detection area, $d_m$ is the distance between the LED transmitter and user $m$, $T_s(\phi_m)$ is the gain of the filter, $g(\phi_m)$ denotes the concentrator gain and is given by $g(\phi_m) = n^2/\sin^2\Phi_m$ with $n$ being the internal refractive index. $R_0(\varphi_m)$ is the Lambertian radiant intensity and given as $R_0(\varphi_m) = [(m+1)/2\pi]\cos^m\varphi_m$, where $m$ is related to the transmitter semiangle at half power and specified as $m = \ln 2/\ln(\cos\phi_{1/2})$. In the simulation setting, we fix the value of $\varphi_{1/2}$ to $60^o$.

Denote by $p_m$ and $s_m$ the transmit allocated power and the message for user $m$, respectively. The composite signal at the transmitter can be expressed, as follows:

$$s = \sum_{m=1}^{M} \sqrt{p_m} s_m + A,$$

where A is a direct current (DC) offset/bias added to ensure the positive instantaneous intensity of the transmitted signal. To maintain non-negativity of the transmitted signal $s$, the following constraint has to be satisfied:

$$\sum_{m=1}^{M} \sqrt{p_m} \leq A/\delta, \quad (1)$$

where $\delta$ is a coefficient, which is determined by modulation order of the pulse amplitude modulation (PAM) employed in the network. While the LED transmitter can transmit at a relatively high power; however, the transmit power should be limited due to considerations of available power consumption of the transmitter and eye safety [7]. To consider eye safety, the transmitted optical intensity should be limited by a peak optical intensity $B$ [3], as the following constraint

$$\sum_{m=1}^{M} \sqrt{p_m} \leq (B-A)/\delta. \quad (2)$$

Let $U_{\max} = \min\{A/\delta, (B-A)/\delta\}$, the constraints (1) and (2) is equivalent to

$$\sum_{m=1}^{M} \sqrt{p_m} \leq U_{\max}.$$

Let $g_m$ be the channel gain between the transmitter and user $m$ and channel gains of all users are sorted in an ascending order, i.e., $|h_1| < ... < |h_m| < ... < |h_M|$. The total bandwidth of network is normalized to unity. Hence, the received signal at user $m$ is given as

$$y_m = h_m \sum_{m=1}^{M} \sqrt{p_m} s_m + z_n,$$

where the DC-offset is already removed, $z_n$ is the additive white Gaussian noise, which is assumed to be i.i.d. and the noise variance is $n_0$. For NOMA, the user $m$ implements the SCI technique to decode the signal of users with lower channel gain; therefore, the SINR at user $m$ to decode signals from users with lower channel gain is

$$\gamma_m = \frac{g_m p_m}{n_0 + g_m \sum_{i=m+1}^{M} p_i},$$

where $g_m = |h_m|^2$. Correspondingly, the achievable rate at user $m$ is $R_m = \log(1+\gamma_m)$ (in this paper, $\log(x)$ returns the natural logarithm of $x$ that is $\ln(x)$).

When the user $m$ achieves the rate $R_m$, its utility is $U_\alpha(R_m)$, where $\alpha$ is the fairness index and $U_\alpha()$ is the $\alpha$ utility function. The utility function $U_\alpha()$ has been well studied in [8] and applied in many problems in wired as well as wireless networks. In this paper, we consider $\alpha = 2$, i.e., harmonic utility. Other values of the fairness index $\alpha$ will be studied in our future work. As the result, the optimization (OPT) problem is formulated as

$$\min_p \sum_{m=1}^{M} \left(\log\left(1 + \frac{g_m p_m}{n_0 + g_m \sum_{i=m+1}^{M} p_i}\right)\right)^{-1}$$

$$s.t. \quad p_m \geq 0, \forall m$$

$$\sum_{m=1}^{M} p_m \leq P_{\max},$$

$$\sum_{m=1}^{M} \sqrt{p_m} \leq U_{\max},$$

where $P_{\max}$ is the maximal transmit power of the transmitter. In our OPT problem, the first constraint guarantees that the transmit power allocated to user $m$ is always non-negative, the second constraint makes sure that the total power allocated to all users does not exceed the maximal value, and the third constraint is to meet the characteristics of power control schemes in VLC systems [3]. In general, the OPT problem is not a convex problem, it is therefore difficult to obtain the optimal solution. In the next Section, we will show that the OPT problem can be equivalently transformed into a convex problem.

## 3. Convexity of the optimization problem

Let us introduce an auxiliary variables $y_m > 0$ satisfying the following constraints

$$\frac{1}{y_m} \leq \log\left(1 + \frac{g_m p_m}{n_0 + g_m \sum_{i=m+1}^{M} p_i}\right)$$

and auxiliary variables $\rho_m = \log(p_m)$. The above inequality is equivalent to

$$\log\left(\frac{n_0}{g_m} e^{-\rho_m} + \sum_{i=m+1}^{M} e^{\rho_i - \rho_m}\right) + \log\left(e^{1/y_m} - 1\right) \leq 0.$$

The OPT problem can be rewritten, as follows:

$$\min_{\{y,\rho\}} \sum_{m=1}^{M} y_m$$

$$s.t. \quad \log\left(\frac{n_0}{g_m} e^{-\rho_m} + \sum_{i=m+1}^{M} e^{\rho_i - \rho_m}\right) + \log\left(e^{1/y_m} - 1\right) \leq 0,$$

$$\sum_{m=1}^{M} e^{\rho_m} \leq P_{\max},$$

$$\sum_{m=1}^{M} e^{\frac{\rho_m}{2}} \leq U_{\max}.$$

It is easy to check that the first part and second part of the first constraint in the above optimization problem are both convex. In addition, the second and third constraints are also convex. As a consequence, the optimal solution to the transformed problem can be obtain by utilizing off-the-shelf convex solvers, for example, interior-point methods [9]. The optimal solution can also be obtained by exploiting the convexity property of the optimization problem and applying the duality approach, it is however not presented here due to the limited space of this paper. After that, we transform the optimal solution back to the original space by $p_m = \exp(\rho_m)$.

## 4. Numerical Simulation

In this section, we provide numerical simulation results to verify the proposed algorithm. Consider a VLC network with a LED transmitter and 20 users, which are uniformly distributed in a coverage of $10\,\text{m} \times 10\,\text{m} \times 3\,\text{m}$. The other simulation parameters are similar to the parameters in [3]. Specifically, the maximal transmit power $P_{\max} = [8:2:20]$ mW, the DC-offset $A = 20$, the peak optical intensity $B = 30$, and the noise power is $n_0 = -104$ dBm.

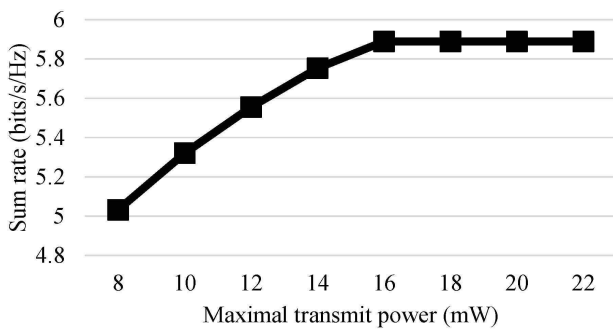

Fig. 1: sum rate of all users under different maximal transmit powers

In our experiment, we show that the sum rate of all users increases with the maximal transmit power of the LED transmitter. However, the sum rate does not increase after a threshold value, i.e., $P_{\max} = 16$ mW. This is reasonable since increasing the transmit power allocated to users improves their sum rate; each user also suffers more interference from other users. It is also observed that at the typical value $P_{\max} = 20$ mW, the sum rate reaches the saturation value.

## 5. Conclusion

We proposed a NOMA-based power control scheme for harmonic-rate utility in VLC systems. The formulated non-convex optimization problem was transformed equivalently into a convex problem, which is solved by the interior-point method to obtain the optimal solution. Simulation result showed performance of the proposed algorithm.

Some extensions can be stemmed from our work. For example, consider other cases of the fairness index. A joint beamforming and power control of NOMA-based visible light communication is highly promising. Moreover, power allocation to guarantee of QoS is obviously needed.

## Acknowledgement

This work was supported by the National Research Foundation of Korea (NRF) grant funded by the Korea government (MSIT) (NRF-2017R1A2A2A05000995). *Dr. CS Hong is the corresponding author.

## References
[1] H. Q. Nguyen *et al.*, "A MATLAB-based simulation program for indoor visible light communication system,” *2010 7th International Symposium on Communication Systems, Networks & Digital Signal Processing*, Newcastle upon Tyne, 2010, pp. 537-541.
[2] H. Marshoud *et al.*, "Non-Orthogonal Multiple Access for Visible Light Communications,” *IEEE Photonics Technology Letters*, vol. 28, no. 1, pp. 51-54, Jan. 2016.
[3] Z. Yang *et al.*, "Fair Non-Orthogonal Multiple Access for Visible Light Communication Downlinks," *IEEE Wireless Communications Letters*, vol. 6, no. 1, pp. 66-69, Feb. 2017.
[4] Q.-V. Pham and W.-J. Hwang, "alpha-fairness resource allocation for non-orthogonal multiple access systems," vol. 12, no. 2, pp. 179-183, Jan. 2018.
[5] L. Yin *et al.*, "On the performance of non-orthogonal multiple access in visible light communication," *2015 IEEE 26th Annual International Symposium on Personal, Indoor, and Mobile Radio Communications*, Hong Kong, 2015, pp. 1354-1359.
[6] X. Zhang *et al.*, "User Grouping and Power Allocation for NOMA Visible Light Communication Multi-Cell Networks," *IEEE Communications Letters*, vol. 21, no. 4, pp. 777-780, Apr. 2017.
[7] J. M. Kahn and J. R. Barry, "Wireless infrared communications,” *Proceedings of the IEEE*, vol. 85, no. 2, pp. 265-298, Feb. 1997.
[8] J. Mo and J. Walrand, "Fair end-to-end window-based congestion control," *IEEE/ACM Transactions on Networking*, vol. 8, no. 5, pp. 556–567, Oct. 2000.
[9] S. Boyd and L. Vandenberghe, *Convex Optimization*. Cambridge University Press, 2004.